\documentstyle[prl,aps,epsf,floats]{revtex}

\def\Abar{\overline{A}}
\def\Bbar{\overline{B}}

\begin{document}

\twocolumn[
\hsize\textwidth\columnwidth\hsize
\csname@twocolumnfalse\endcsname

\title{Measuring $\gamma$ Cleanly with CP-Tagged $B_s$ and
$B_d$ Decays}
\author{Adam F. Falk and Alexey A. Petrov\\[4pt]}
\address{\tighten{\it
Department of Physics and Astronomy, 
The Johns Hopkins University,\\
3400 North Charles Street, 
Baltimore, Maryland 21218}\\[4pt]
hep-ph/0003321\\[4pt]
March, 2000}

\maketitle

\tighten{
\begin{abstract}

We propose a new method for measuring the CKM phase
$\gamma$ using the partial rates for CP-tagged $B_s$
decays.  Such an experiment could be performed at a very
high luminosity symmetric $e^+e^-$ collider operating at the
$\Upsilon(5S)$ resonance, where the $B_s\Bbar_s$ state is
produced in a state of definite CP.  We also discuss
CP-tagging in the $B_d$ system at the $\Upsilon(4S)$, where a
time-dependent analysis is required to compensate for the
anticipated large CP violation in $B_d-\Bbar_d$ mixing.
\end{abstract}
}
\vspace{0.2in}

]\narrowtext

\newpage

The accurate determination of the Unitarity Triangle of the
Cabibbo-Kobayashi-Maskawa (CKM) quark mixing matrix  is one of
the most important problems of experimental $B$ physics. 
Reasonably precise measurements of some of its parameters --
three sides, determined by $|V_{ub}|$, $|V_{cb}|$ and
$|V_{td}|$, and one angle $\sin2\beta$ -- will soon be
performed at the $B$ Factories operating at the $\Upsilon(4S)$
and at Run II of the Tevatron.  Although this is enough to fix
the triangle up to discrete ambiguities, one really wants to
overconstrain the system and thereby be sensitive to
deviations from the CKM description of flavor-changing
processes.  In view of this goal, it is important to measure
the angles $\alpha$ and
$\gamma$ as well.

The situation with these other angles is more problematic. A
number of methods have been proposed to measure or constrain
$\alpha$ and $\gamma$, but unfortunately they each suffer to
some degree from either  theoretical or experimental
difficulties~\cite{Harrison:1998yr}.  In what follows we
investigate a new proposal to constrain  $\gamma$ in the
decays of two $B_s$ mesons produced in a coherent state.  This
can be achieved if the pair comes from the decay of a $b\bar
b$ meson such as the $\Upsilon (5S)$.  In this case one has
not only the option of tagging the flavor of the
initial $B_s$, but the alternative of tagging it as an
eigenstate of CP.  The advantage of the $B_s$ in this regard
is that CP violation in its mixing is small in the Standard
Model.  We will also discuss an analogous proposal
for the $B_d$ system at the $\Upsilon(4S)$, where large CP
violation in mixing complicates the situation. 

The CKM angle $\gamma$ is defined to be
\begin{equation}
  \gamma = \arg \left [ - \frac{V_{ud}V_{ub}^*}{V_{cd}V_{cb}^*}
  \right].
\end{equation}
Here we consider the possibility of extracting $\gamma$ from
the $B_s$ decays to the final states $D_s^\pm K^\mp$ (or
the analogous $D{}_s^{(*)}K^{(*)}$
combinations).  The fact that
$D_s^- K^+$ and $D_s^+ K^-$ can be reached from both the $B_s$
and its CP conjugate $\Bbar_s$ already has been exploited in a
proposal to extract $\gamma$ from a time-dependent
study~\cite{Aleksan:1992nh}.  The two transition amplitudes
have similar magnitudes, $|A(B_s\to D_s^- K^+)| \sim
|A(\Bbar_s\to D_s^- K^+)| \sim \lambda^3$, where
$\lambda=\sin\theta_C\simeq0.22$ is the small parameter which
controls the hierarchy of the CKM matrix.  Hence triangles
built from these amplitudes need not suffer from being
``squashed''.  Given sufficient statistics, the time-dependent
analysis eventually should yield $\sin\gamma$ at some level of
accuracy.  

By contrast, our proposal allows one to measure
$\sin\gamma$ using branching ratios only, with no need to
determine the time at which the decay occurs.  What will be
necessary, instead, is to measure not only flavor-tagged
but also CP-tagged $B_s$ decays.

We begin by defining amplitudes for $B_s$ and $\Bbar_s$ decay
to the final state $D_s^-K^+$,
\begin{eqnarray} \label{ampl1}
  A_1 &=& A(B_s \to D_s^-K^+) = a_1 e^{i \delta_1}\,,
  \nonumber \\
  A_2 &=& A(\Bbar_s \to D_s^-K^+) 
  = a_2 e^{-i \gamma} e^{i \delta_2}\,.
\end{eqnarray}
The amplitude $A_1$ arises from the quark transition $\bar
b\to \bar cu\bar s$ and is real (in the Wolfenstein
parameterization), while $A_2$ arises from $b\to u\bar cs$
and carries the relative weak phase $e^{-i\gamma}$.  There
are no penguin contributions.  The amplitudes $A_i$
also have strong phases $e^{i\delta_i}$. The CP conjugated
amplitudes are given by
\begin{eqnarray} \label{ampl1bar}
  \Abar_1 &=& A(\Bbar_s \to D_s^+ K^-) = a_1 e^{i
  \delta_1}e^{-2i\xi}\,,
  \nonumber \\
  \Abar_2 &=& A(B_s \to D_s^+ K^-) 
  = a_2 e^{i \gamma} e^{i \delta_2}e^{2i\xi}\,,
\end{eqnarray}
where the phase $\xi$ depends on the convention for CP
transformations of the $B_s$ states,
\begin{equation}
  {\rm CP}|B_s\rangle=e^{2i\xi}|\overline{B}_s\rangle\,,\quad
  {\rm CP}|\overline{B}_s\rangle=e^{-2i\xi}|B_s\rangle\,.
\end{equation}
Any physical observable must be independent of $\xi$.  We also
define a set of  amplitudes for the CP eigenstates of the
$B_s$ meson,
\begin{equation}\label{cpstates}
  | B_s^\pm \rangle = \frac{1}{\sqrt{2}} \Big[| B_s\rangle
  \pm  e^{2i \xi} | \overline{B}_s \rangle \Big],
\end{equation} 
to decay into the same $D_s K$ final states,  
\begin{eqnarray} \label{amplcp}
  A_\pm &=& A(B_s^\pm \to D_s^- K^+)\,,
  \nonumber \\
  \Abar_\pm &=& A(B_s^\pm \to D_s^+ K^-)\,.
\end{eqnarray}
These amplitudes satisfy simple triangle relations,
\begin{eqnarray}\label{triangrels1}
  \sqrt2\,A_\pm&=&A_1\pm e^{2i\xi}A_2=(a_1\pm a_2e^{-i\gamma}
  e^{i\delta})e^{i\delta_1}\,,\nonumber\\
  \sqrt2\,\Abar_\pm&=&\Abar_2\pm e^{2i\xi}\Abar_1=\pm(a_1\pm
  a_2e^{i\gamma}e^{i\delta})e^{i\delta_1}\,,
\end{eqnarray}
where $\delta=\delta_2-\delta_1+2\xi$ is the
convention-independent (and observable) strong phase
difference.  It is clear that any construction which is
insensitive to $\delta$ will also be insensitive to the
unphysical phase $\xi$.  It is also clear that by changing
$\xi$, it is possible to take $B_s^\pm$ to be a linear
combination of $B_s$ and $\Bbar_s$ with any relative
phase.  We will derive a relation for $\sin2\gamma$
involving the magnitudes of the amplitudes $A_i$ and
$\Abar_i$.  From the freedom to choose $\xi$ in
Eq.~(\ref{triangrels1}), it is clear that the CP-even and
CP-odd amplitudes will yield triangle relations which contain
the same information about $\gamma$.  

\begin{figure}
\centering
\epsfysize 1.5in
\centerline{
\epsfbox{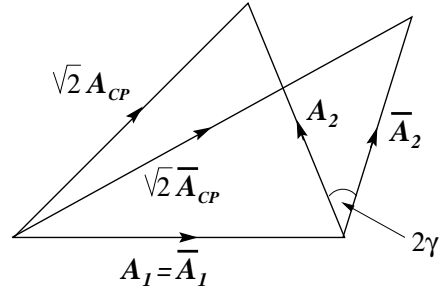}}
\caption{Amplitude triangle relations for $B_s \to D_s K$.}
\label{fig1}
\end{figure}

For any CP eigenstate $B_s^{\rm CP}$, then, it is
possible to choose $\xi$ so that
\begin{eqnarray}\label{triangrels}
  A_{\rm CP} &=& A(B_s^{\rm CP}\to
  D_s^-K^+)=(A_1+A_2)/\sqrt2\,,\nonumber \\
  \Abar_{\rm CP} &=& A(B_s^{\rm CP} \to D_s^+
  K^-)=(\Abar_1+\Abar_2)/\sqrt2\,.
\end{eqnarray}
Without loss of generality, we also choose a convention in
which $\delta_1=0$, in which case the triangle relations are
very simple.  They are illustrated graphically in
Fig.~\ref{fig1}, where the amplitudes may be interpreted as
vectors in the complex plane.  As drawn, the angle between
$A_2$ and $\Abar_2$ is $2\gamma$.  For an analytical solution,
it is convenient to define
\begin{eqnarray}\label{alphas}
  \alpha&=&{2|A_{\rm CP}|^2-|A_1|^2-|A_2|^2\over
  2|A_1||A_2|}\,,\nonumber\\
  \overline\alpha&=&{2|\Abar_{\rm CP}|^2-|\Abar_1|^2
  -|\Abar_2|^2\over2|\Abar_1||\Abar_2|}\,,
\end{eqnarray}
in terms of which we find
\begin{equation}\label{sin2gamma}
  \sin2\gamma=\pm\left(\alpha\sqrt{1-\overline\alpha^2}
  -\overline\alpha\sqrt{1-\alpha^2}\right).
\end{equation}
The determination of $\gamma$ itself then has an eightfold
ambiguity.  This construction is quite analogous to that of
Ref.~\cite{Gronau:1991dp}, in which the extraction of
$\sin\gamma$ from $B_d$ decay to a CP eigenstate $D_{\rm CP}$
was studied.  (However, the triangles are ``squashed'' in that
analysis.)  Note that if $\Delta\Gamma_s/\Gamma_s$ is
significant, the squared amplitudes $|A_i|^2$ and
$|\Abar_i|^2$ are proportional to partial rates, {\it e.g.},
\begin{equation}
  |A_1|^2\propto \Gamma(B_s\to D^-_sK^+)\,,
\end{equation}
rather than to branching ratios.

The measurement of $|A_{\rm CP}|$ requires that one tag the initial
$B_s$ state as a CP eigenstate $B_s^{\rm CP}$.  This is possible if
the $B_s$ is produced in the decay $\Upsilon(5S)\to
B_s\Bbar_s$.  Since the $\Upsilon(5S)$ is a CP even state
and the $B_s$ and $\Bbar_s$ are emitted in a relative
$p$-wave, the CP eigenvalues of the $B_s/\Bbar_s$ mixtures are
anti-correlated.  Hence if the ``tagging'' $B_s$ decays to a
CP eigenstate such as $D_s^+D_s^-$, then the other $B_s$ is
constrained to be a CP eigenstate as well.  It is crucial that
the tagging decay be one in which direct CP violation is
expected to be small.  Tagging modes with spin one particles,
such as $B_s^{\rm CP}\to\psi\phi$ and $D_s^*\overline
D{}^*_s$, can be used if an angular analysis is performed to
select a final state of definite CP.  So long as the
amplitudes $|A_{\rm CP}|$ and $|\Abar_{\rm CP}|$ are measured
with the same CP tagging mode on the opposite side, it is
unimportant whether a CP even or CP odd tag is employed.

This simple method of CP tagging relies on the Standard Model
expectation that CP violation in $B_s$ mixing is not
significant.  As $B_s$ mixing is generated dominantly by
$t-W$ box diagrams, CP violating effects are proportional
to $\sin2\beta_s$, where
\begin{equation}
  \beta_s=\arg\left[-\frac{V_{ts}V_{tb}^*}{V_{cs}V_{cb}^*}
  \right]\sim\lambda^2
\end{equation}
is small.  Furthermore, we assume that CP
violation in $B_s$ mixing will have been constrained
experimentally by the time the analysis proposed here is
performed.

The method is not appropriate to a hadronic production
environment such as the Tevatron or the LHC, since in this
case the $B_s$ and $\Bbar_s$ do not arise from an initial CP
eigenstate.  Nor, of course, can this analysis be performed at
the $B$ Factories as presently configured to operate at the
$\Upsilon(4S)$.  To our knowledge, this is the first proposal
for a clean measurement of a CKM phase which is unique to an
$e^+e^-$ collider operating at the $\Upsilon(5S)$.

We now make a crude estimate of the number of $B_s \Bbar_s$
pairs required to measure $\sin\gamma$ with a precision of
$0.1$, for which approximately $10^2$ reconstructed events
would be needed.   To be concrete, we take the tagging mode
$B_s^{\rm CP}\to D_s^+ D_s^-$.  With order-of-magnitude
estimates of the relevant branching ratios,
${\cal B}(B_s^{\rm CP} \to D_s^+ D_s^-) \sim 10^{-2}$ and
${\cal B}(B_s^{\rm CP} \to D_s K) \sim 2 \times 10^{-4}$, and
assuming that the $D_s$ can be reconstructed efficiently
by combining a number of decay modes, we find a combined
CP-tagged branching fraction ${\cal B}_{\rm tot}\sim10^{-6}$. 
Hence approximately $10^8$ $B_s \Bbar_s$ events would be
needed for this measurement.  

The decays of  the $\Upsilon(5S)$ to $B_s$ flavored mesons
produce primarily the combination $B{}_s^*\Bbar{}^*_s$, as
well as $B{}_s^*\Bbar_s$, $B_s\Bbar{}^*_s$ and $B_s\Bbar_s$. 
The relative rates have been computed in a variety
of models, yielding the estimates~\cite{production}
\begin{equation}
  \sigma(B_s\Bbar_s)/\sigma(B{}_s^*\Bbar{}^*_s)
  \approx 0.1-0.2
\end{equation}
and
\begin{equation}
  \sigma(B{}_s^*\Bbar_s+B_s\Bbar{}^*_s)/
  \sigma(B{}_s^*\Bbar{}^*_s)\approx 0.05-0.5\,.
\end{equation}
A $B{}_s^*$ produced in this way decays to a $B_s$ and a
very soft photon, so the other combinations will also be
seen as $B_s\Bbar_s$.  In fully reconstructed $B_s$
decays the combinations can be separated by measuring the
boost of the $B_s$~\cite{Harrison:1998yr}.  From the ratio
$\sigma(B{}_s^*\Bbar{}^*_s)/ \sigma(\Upsilon(4S))\simeq0.1$ of
production cross sections and the fact that an $e^+e^-$
collider with luminosity ${\cal L}=10^{33}\,{\rm cm}^{-2}{\rm
sec}^{-1}$ produces $3.6\times10^7$ $\Upsilon(4S)$ events per
year~\cite{Harrison:1998yr}, we see that $10^2$ CP-tagged
$B_s\Bbar_s$ events, decaying in this mode, are within the
reach of a $B$ Factory upgraded to operate at
${\cal L}\approx10^{35}\,{\rm cm}^{-2}{\rm sec}^{-1}$.   Since
it is not necessary to measure the time-dependence of the
decay, the experiment could be performed in a future high
luminosity run of the Cornell $e^+e^-$ storage ring. 

Moreover, there are ways to increase substantially the
number of usable events.  First, one may repeat the analysis
with the final states $D_s^*K$, $D_sK^*$ and $D_s^*K^*$. 
Note that no angular analysis is necessary here, since one is
not isolating a CP eigenstate on the decay side.  Second, one
can add additional CP-tagging modes such as $B_s^{\rm
CP}\to\psi\phi$ or $D_s^*\overline D{}^*_s$.  Although in this
case an angular analysis would be required to separate final
states of definite CP, studies in the $B_d$ system indicate
that this can be done without a large cost in tagging
efficiency~\cite{Harrison:1998yr}.  This gain in efficiency
will be offset in part by the cost of fully reconstructing the
$D_s$ states, a penalty which we have not explicitly included.

Finally, it also may be possible to use the $B{}_s^*\Bbar_s$
and $B_s\Bbar{}^*_s$ combinations for CP tagging.  Parity
conservation requires that the pair be produced in a
relative $p$-wave.  Therefore the initial state is of the form
\begin{equation}
  {1\over\sqrt2}\big[B_s^+B_s^{*-}+B_s^-B_s^{*+}\big]\,,
\end{equation}
where $B_s^{*\pm}$ are the CP eigenstate mixtures of
$B{}_s^*$ and $\Bbar{}_s^*$, in analogy with the $B_s$
combinations (\ref{cpstates}).  After the transition
$B{}_s^*\to B_s\gamma$, in which the magnetic photon carries
${\rm CP}=-1$, the CP eigenvalues of the $B_s/\Bbar_s$
mixtures on the two sides are correlated (rather than
anti-correlated, as in direct $\Upsilon(5S)\to
B_s\Bbar_s$).  As we have shown, our analysis is equivalent
for correlated and anti-correlated states.  Unfortunately, it
is not possible to CP tag using the dominant
$B{}_s^*\Bbar{}_s^*$ combination, since the total spin quantum
number is not fixed at production.

Taken together, the use of additional modes on the tagging
and decay sides and of the $B{}_s^*\Bbar_s$ and
$B_s\Bbar{}^*_s$ initial states should allow one to relax
considerably the luminosity requirement estimated above. 
Alternatively, for a given integrated luminosity these
enhancements would allow a statistically more precise
measurement of $\sin\gamma$.  Generally speaking, we believe
that our proposal is feasible within many of the scenarios
under discussion for future luminosity upgrades of  the $B$
Factories now operating at the
$\Upsilon(4S)$. 

As shown above, the measurement of $\gamma$ from this analysis
is insensitive to the strong phase difference $\delta$ between
the amplitudes $A_1$ and $A_2$.  In fact, $\delta$ could be
extracted simultaneously with $\gamma$ from the amplitude
triangles shown in Fig.~\ref{fig1}.  Nevertheless, it is
useful to have some idea of whether $\delta$ may be expected
to be large.  It is clear that elastic rescattering of the
$D_sK$ final state will be the same for $B_s$ and $\Bbar_s$
transitions and will not lead to nonzero $\delta$.  Instead,
what is needed to generate
$\delta\ne0$ is rescattering through an intermediate state $f$
which is produced differently by $B_s$ and
$\Bbar_s$.  Then we may write
\begin{eqnarray}
  \frac{A(B_s \to D_s^- K^+)}{A(\Bbar_s \to D_s^- K^+)} &=& 
  \frac{A_1^{\rm dir} + \epsilon_f A_1^f + \dots}
  {A_2^{\rm dir} + \epsilon_f A_2^f + \dots}
  \nonumber \\
  &\simeq &
  \frac{A_1^{\rm dir}}{A_2^{\rm dir}}\cdot
  \frac{1 + \epsilon_f A_1^f/A_1^{\rm dir}}
  {1 + \epsilon_f A_2^f/A_2^{\rm dir}}\,,
\end{eqnarray}
where $A_i^{\rm dir}$ are the amplitudes for the direct
production of $D_s^- K^+$, $A_i^f$ are the amplitudes for the
production of the intermediate state $f$, and $\epsilon_f$ is
the amplitude for the rescattering $f\to D_s^- K^+$.  While
$A_1^{\rm dir}$ and $A_2^{\rm dir}$ have the same strong phase,
a strong phase difference can be generated if $A_1^f/A_1^{\rm
dir}\ne A_2^f/A_2^{\rm dir}$.  For a similar discussion in the
context of $D$ decays, see Ref.~\cite{Falk:1999ts}.

To estimate the size of $\delta$ which could be generated, we
consider a model in which $f$ is a two body intermediate
state.  We employ the formalism of Ref.~\cite{Falk:1998wc},
based on Regge phenomenology and naive factorization of the
$B_s$ decay matrix elements.  With $f=D_s^* K^*$ and
rescattering to $D_sK$ via exchange of the Pomeron and $\phi$
trajectories, we find
\begin{equation} \label{phase}
  \delta < 5^\circ.
\end{equation}
Although our estimate is extremely model-dependent, it does
provide some evidence that this mechanism is unlikely to
produce a large value of $\delta$.  We note that a perturbative
factorization  formalism is not applicable to the decay $B_s
\to D_sK$.

Finally, we turn to the issue of CP tagging in $B_d$ decays. 
Here the situation is complicated by the fact that the
Standard Model predicts large CP violation in $B_d$ mixing. 
Therefore a state which is tagged at time $t=0$ as being in a
CP eigenstate will evolve by time $t$ into a linear combination
$B_d^\pm(t)$ of the CP even ($B_d^+$) and CP odd ($B_d^-$)
states.  The evolution is given by
\begin{equation}
  B_d^\pm(t)=e^{-i(M_B+\Gamma/2)t}
  \left(a_\pm(t)B_d^\pm+b_\pm(t)B_d^\mp\right)\,,
\end{equation}
where
\begin{eqnarray}\label{evolution}
  a_\pm(t)&=&\cos\left({\Delta m_dt/2}\right)\pm
  i\cos2\beta\sin\left({\Delta m_dt/2}\right)\,,\nonumber\\
  b_\pm(t)&=&\mp\sin2\beta\sin\left({\Delta m_dt/2}\right)\,,
\end{eqnarray}
and $\Delta m_d$ is the mass splitting between $B_H$ and
$B_L$.  Here
\begin{equation}
  \beta=\arg\left[-\frac{V_{td}V_{tb}^*}{V_{cd}V_{cb}^*}
  \right]  
\end{equation}
is an angle which is expected to be large.  (If $B_d$
mixing receives a significant contribution from new physics,
then the CP violating phase ``$\sin2\beta$'' extracted from the
asymmetry in $B_d\to J/\psi K_S$ is actually what governs the
time evolution (\ref{evolution}).)  Note that in the CP
conserving limit
$\sin2\beta\to0$, we have
$a_\pm(t)\to\exp(\pm i\Delta m_dt/2)$ and $b_\pm(t)\to0$, so
the masses of the CP eigenstates are shifted to
$M_B\pm{1\over2}\Delta m_d$ but the states do not mix.  

In analogy with the $B_s$ case, we define amplitudes for a
$B_d$, tagged at $t=0$ as a CP eigenstate $B_d^\pm$, to decay
into the final states $D^\pm\pi^\mp$ (or $D^*\pi^\mp$ or
$D^{(*)}\rho^\mp$) at time~$t$,
\begin{eqnarray}
  A_\pm(t) &=& A(B_d^\pm(t) \to D^-\pi^+)\,,
  \nonumber \\
  \Abar_\pm(t) &=& A(B_d^\pm(t) \to D^+\pi^-)\,.
\end{eqnarray}
The triangle relations analogous to Eq.~(\ref{triangrels})
then take a form which depends on $t$,
\begin{eqnarray}\label{Bdtriangs}
  \sqrt2\,\big|A_\pm(t)\big| &=&
  \big|r_\mp(t)A_1+r_\pm(t)A_2\big|\,,
  \nonumber \\
  \sqrt2\,\big|\Abar_\pm(t)\big| &=&
  \big|r_\pm(t)\Abar_1+r_\mp(t)\Abar_2\big|\,,
\end{eqnarray}
where here the amplitudes $A_i$ and $\Abar_i$ are defined as
in Eqs.~(\ref{ampl1}) and ({\ref{ampl1bar}) but for
$B_d\to D\pi$, and
\begin{equation}
  r_\pm(t)=\big[1\pm\sin2\beta\sin\Delta m_dt\big]^{1/2}.
\end{equation}
One may extract $\gamma$ by fixing a value of $t$ and then
constructing the amplitude triangle with the sides scaled by
$r_\pm(t)$ as in Eq.~(\ref{Bdtriangs}).  The expressions for
$\alpha$ and $\overline\alpha$ are modified to
\begin{eqnarray}\label{alphast}
  \alpha_\pm(t)&=&{2|A_\pm(t)|^2
  -r_\mp^2(t)|A_1|^2-r_\pm^2(t)|A_2|^2
  \over2r_+(t)r_-(t)|A_1||A_2|}\,,\nonumber\\
  \overline\alpha_\pm(t)&=&{2|\Abar_\pm(t)|^2
  -r_\pm^2(t)|\Abar_1|^2-r_\mp^2(t)|\Abar_2|^2
  \over2r_+(t)r_-(t)|\Abar_1||\Abar_2|}\,.
\end{eqnarray}
The solution (\ref{sin2gamma}) for $\sin2\gamma$, written in
terms of $\alpha_\pm(t)$ and $\overline\alpha_\pm(t)$, is
independent of $t$ by construction.  Note that for this
time-dependent analysis, the decays of the CP even and CP odd
eigenstates are not equivalent.

The procedure may be repeated to give an independent
measurement of $\gamma$ for each bin in $t$.  Writing the
amplitude triangles in the form (\ref{Bdtriangs}), a cosmetic
change which makes obvious the generalization of $\alpha$ and
$\overline\alpha$ to $\alpha_\pm(t)$ and
$\overline\alpha_\pm(t)$, requires a $t$-dependent choice of
the CP transformation phase $\xi$.  This is legitimate, since
in Eq.~(\ref{Bdtriangs}) one combines amplitudes only at a
fixed value of $t$.

The necessity of determining the decay time $t$ means that such
a measurement of $\gamma$ in the $B_d$ system would have to
be performed at an asymmetric $B$ Factory operating at the
$\Upsilon(4S)$.  Although in principle the analysis could be
performed by the BaBar or BELLE Collaborations, there are
several difficulties.  First, an accurate independent
determination of $\sin2\beta$ must be available.  Second, it
is necessary to collect sufficient statistics to construct the
amplitude triangles for individual bins in $t$.  Third, in the
case of $B_d\to D^\pm\pi^\mp$, $D^*\pi^\mp$ or
$D^{(*)}\rho^\mp$ the amplitude triangles are
``squashed'', with one side shorter than the other two by a
factor of order $\lambda^2$.  (They would not be squashed,
however, if the analysis were performed instead for a mode
such as $B_d\to D^{(*)}K_S$.) 

In summary, we have presented a new approach to extracting
the CKM angle $\gamma$, employing an analysis which
depends on tagging an initial $B_s$ or $B_d$ as a CP
eigenstate.  This theoretically clean method is free from
dependence on unknown strong phases.  In the $B_s$ case, the
analysis is unique to an experiment performed at a very high
luminosity $e^+e^-$ collider operating at the $\Upsilon(5S)$
resonance.  While there are no definite plans to upgrade any
of the existing symmetric or asymmetric $B$ Factories to
operate in this mode, we hope that the proposal outlined here
will help rekindle interest in this possibility.

\smallskip

It is a pleasure to thank Yossi Nir and Helen Quinn for helpful
correspondence.  Support was provided by the National Science
Foundation under Grant PHY--9404057, by the Department of
Energy under Outstanding Junior Investigator Award
DE--FG02--94ER40869, and by the Research Corporation under
Cottrell Scholar Award CS0362.

\tighten

\end{document}